# Demonstration of Multi-Active Region P-down Green LEDs with High Quantum Efficiency

Sheikh Ifatur Rahman[1], Robert Armitage[2] and Siddharth Rajan[1,3]

[1]*Department of Electrical and Computer Engineering, The Ohio State University, Columbus, Ohio 43210, USA.*
[2]*Lumileds LLC, San Jose, California, 95131, USA.*
[3]*Department of Materials Science and Engineering, The Ohio State University, Columbus, Ohio 43210, USA.*
Authors to whom correspondence should be addressed: rahman.227@osu.edu ; rajan.21@osu.edu

**Abstract:**

Longer wavelength emitters such as green LEDs display a pronounced efficiency drop at higher current densities, resulting in relatively low wall-plug efficiency. Multi-active region approach can improve the wall-plug efficiency significantly and tackle the "green gap" challenge. This work reports multi-active region p-down LEDs with high external efficiency operating entirely in the green wavelength. Devices were developed using p-down topology, where the PN junction is oriented such that electric fields from depletion and built-in polarization dipoles are aligned. Ga-polar multi-active region green LEDs with excellent voltage and EQE scaling, and significantly higher wall-plug efficiency is demonstrated in this work.

The III-nitride material system has gained significant attention for its potential in the development of a wide range of electronic devices, such as light emitting diodes (LEDs), transistors, and detectors,[1-5] but emitters in the longer wavelength ranges, such as green, remain a significant challenge. InGaN/GaN short wavelength (blue/violet) light emitters have been commercialized for lighting, display, and laser applications. The power conversion efficiency for blue/violet LEDs have reached near 80% which exceeds traditional forms of lighting such as incandescent or compact fluorescent lamps while being safe for the environment.[6] In several applications, there is a need to obtain high areal power density from LEDs, while maintaining high efficiency. However, while the peak efficiency values are high, the efficiency of emitters decreases dramatically as the input power increases. This efficiency droop phenomenon impacts the overall power conversion efficiency, and becomes more pronounced at longer wavelengths (> 500 nm).[7] These green (and longer wavelength) emitters have been found to display a pronounced drop in efficiency with current density (efficiency droop), leading to a relatively low wall-plug efficiency (WPE) at the current density needed for typical applications. The drop in external quantum efficiency (EQE) and wall-plug efficiency at higher current density makes it challenging to achieve high power density and high efficiency simultaneously. One approach to potentially improve efficiency or wall-plug performance is to use tunnel junction-based multi-active region LED structures,[8-11] which enable higher optical power output from multiple active regions. The wall-plug efficiency of dual junction LED is expected to be higher than a single junction LED for a given output power density. This is due to the efficiency droop phenomenon in III-Nitride LEDs. To achieve higher power density, single junction LEDs must be operated at higher current densities where the efficiency is lower than the peak efficiency (due to efficiency droop). Since a two-junction LED has two active regions, it requires lower current density to achieve the same power as a single-junction LED. The operation at the lower current density enables the two-junction LED to have higher external quantum efficiency, and thereby higher wall-plug efficiency than a single-junction LED. This has also been discussed in a few previous paper papers from Akyol et al, and Jamal-Eddine et al.[10, 12] The use of multiple active regions essentially enables us to achieve higher output power density *from the same chip area* without sacrificing efficiency.

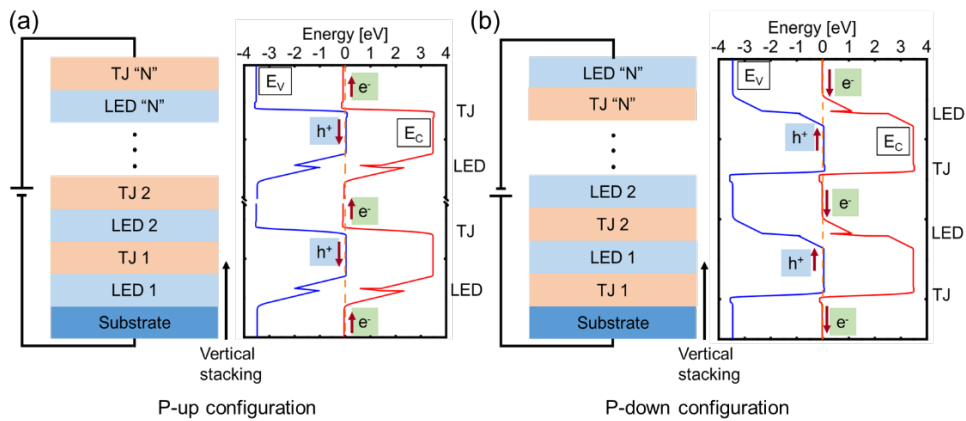

**Figure 1.** (a) Schematic and equilibrium band diagram of a double active region LED structure using tunnel junctions in the conventional p-up configuration (b) Schematic and equilibrium band diagram of a double active region LED in the p-down configuration of LED.

Figure 1 (a) and 1 (b) show the schematic and energy band diagrams for p-up and p-down double active region LED structures using tunnel junctions respectively. Using multiple LEDs in series configuration enables higher power through voltage, rather than current scaling. At a given forward current ($I_F$), the power and EQE for an ideal N-active region structure are higher by a factor 'N', with a voltage drop which is also higher by a factor 'N'. Thus, when using the conventional definition of external quantum efficiency (photons emitted/electrons injected), the external quantum efficiency of such a structure is 'N' times higher than a single-junction LED. The key advantage of this structure is that it provides a way to circumvent the efficiency droop. At any given current, the N-active region LED provides 'N' times higher power density. This allows us to achieve the high efficiency associated with a lower current density, while also emitting higher optical power. This is equivalent *electrically* to N single junction LEDs connected in series, but it requires less area than that approach since all the LEDs are stacked vertically. In this work, we demonstrate InGaN/GaN multi-active region *green* LEDs with near-ideal voltage scaling and high external efficiency.

Our approach leverages the p-down polarization-reversed topology to reverse the relative electric field of the PN junction and polarization fields. In conventional metal-polar (+c oriented) LEDs, where the p-doped layer is on top of the n-doped region, the polarization dipole within the InGaN quantum well opposes the depletion field, and high electrostatic carrier injection barriers are formed on both sides of the well. These barriers block electron and hole injection into the active region, and therefore could cause degradation in the electrical efficiency of the diode. In the p-down case, the polarization dipoles are in the same direction as the depletion field, and therefore, the electrostatic barriers to electron and hole injection at the edge of the quantum wells are reduced. This reduction in electrostatic barriers enables lower turn-on and forward voltage ($V_F$) of operation and possibly improves carrier injection into the active region.[13-18] A major challenge for a p-down structure is that the p-doped layer at the bottom has relatively high contact resistance and sheet resistance due to the low mobility of holes, which leads to current crowding if the p-type contact is made by etching the top n-region and making contact to the bottom p-region.[19-21] An interband tunnel junction below the p-layer solves this problem since the bottom n-region can be contacted and it can efficiently spread current and prevent current crowding. Tunnel junctions were previously demonstrated for current spreading layers without the use of absorbing indium tin oxide (ITO) contacts to p-type GaN,[22, 23] efficient hole injection into multiple-quantum well (MQW) active regions,[24-26] and cascading of multiple LED active regions.[27-30] Previous reports of MBE-grown p-down green LED has shown that p-down LEDs grown along Ga-polar direction with well-engineered TJ design can achieve high electrical efficiency owing to the benefits from lowered electrostatic barrier, yet the demonstrated LED suffered from low quantum efficiency due to the induced excess defects from growth condition of the MBE.[13] Recently, metalorganic chemical vapor deposition-deposited p-down green LEDs using bottom tunnel junctions with high external quantum efficiency were demonstrated.[31] Although the demonstrated p-down LED + TJ device showed higher forward voltage drop due to extra voltage penalty at the TJ (caused from the challenges associated with the high Mg doping and its profile in the MOCVD growth), the extracted p-down LED (standalone) forward voltage drop showed that the diode in reversed polarization topology can exhibit lower voltage drop compared to the conventional p-up structure. In this study, we utilize the reverse polarization topology and demonstrate p-down LEDs comprising with two active regions along the Ga-polar orientation of the crystal, operating entirely in the green wavelength. Figure 1(b) depicts a schematic of the multi-active region p-down LED utilizing tunnel junctions along the equilibrium band diagram for dual active region. The working principle and performance benefits are applicable in the similar manner for p-down LEDs as compared to the p-up configuration. The tunnel junction in this p-down configuration, at the bottom injects hole into the bottom active region under reverse bias (while the LED is forward biased) and the tunnel junction at the middle of two active region simultaneously injects hole to the top active region and injects electron to the bottom active region.

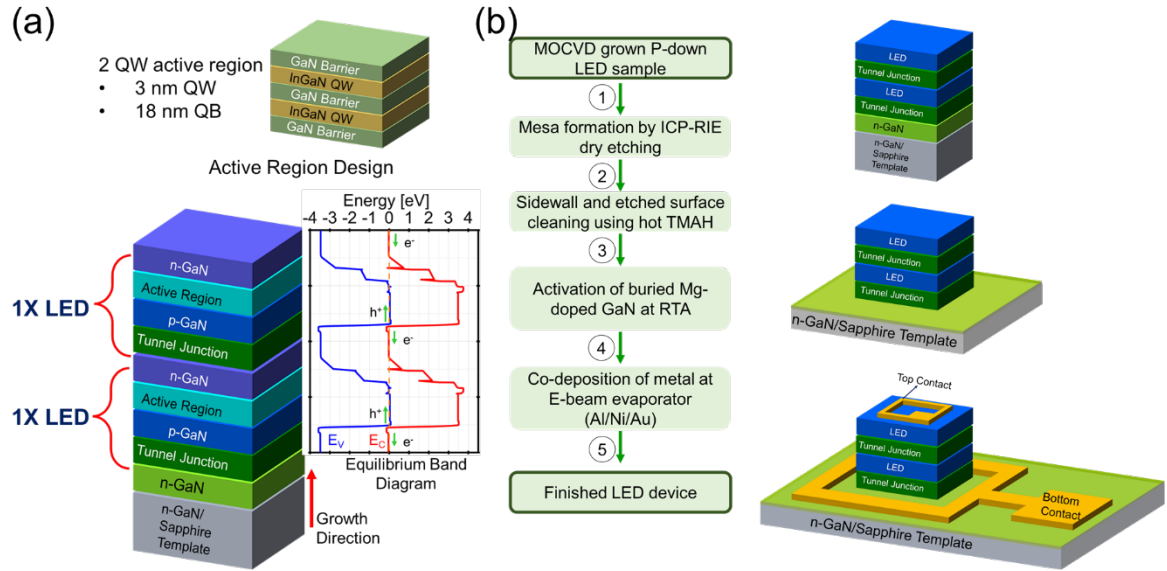

**Figure 2.** (a) Epitaxial structure of the multi-active p-down LED with each active region design and corresponding equilibrium band diagram (b) Process flow for single junction and dual junction p-down LEDs.

The epitaxial structure with the corresponding equilibrium band diagram for the multi junction p-side down LED is shown in figure 2(a). The active region (figure 2(a)) is comprised of two 3 nm InGaN quantum wells sandwiched between 18 nm GaN barriers. The LED epitaxial structures were grown on a n-type GaN on sapphire (non-patterned) template without any surface texturing or encapsulation. The bottom tunnel junction ($n^{++}/p^{++}$) allows the growth of the p-type GaN layer before the active region and replaces the p-type contact with a low-resistance n-type contact and spreading layer. The tunnel junction in the middle allows seamless integration of the second p-down LED at the top. The device fabrication process is shown in figure 2 (b). Device fabrication was carried out with patterning by direct-write optical lithography (Heidelberg MLA150). Mesa isolation of the devices was done using inductively coupled plasma-reactive ion etching (ICP-RIE) using $BCl_3/Cl_2/Ar$ chemistry. The etched surface was then cleaned using $80°C$ hot TMAH to remove any residues on the etch surface after the dry etch. The activation of buried p-type GaN was done through sidewall using rapid thermal annealing. The sidewall activation process of buried p-GaN for similar multi-active region structure has been discussed with more details in previous reports.[28, 32-34] Bottom and top n-type contacts (Al/Ni/Au - 30 nm/30 nm/100 nm) were deposited at the same time using e-beam evaporation reducing extra process steps for both single junction and dual junction LED. This non-alloyed metal electrodes to the n-type GaN at the top and bottom shows completely ohmic behavior. Electrodes on the devices have not been optimized to maximize reflectivity and light extraction. The LED devices were then characterized on-wafer using devices with 90% metal coverage at the top to characterize the electrical and optical properties. Test setup does not collect all the emitted light. The electrical characteristics were measured using Keysight B1500A semiconductor device analyzer. The electroluminescence measurement was performed on-wafer for devices with ring shaped contact on top using an Ocean Insight Modular Spectrometer. The output power (collected from the bottom of the wafer) was measured using a Thorlabs PM100D optical power meter fitted with a S120VC photodiode power sensor. The data reported here obtained from 100 $\mu m$ × 100 $\mu m$ device dimensions.

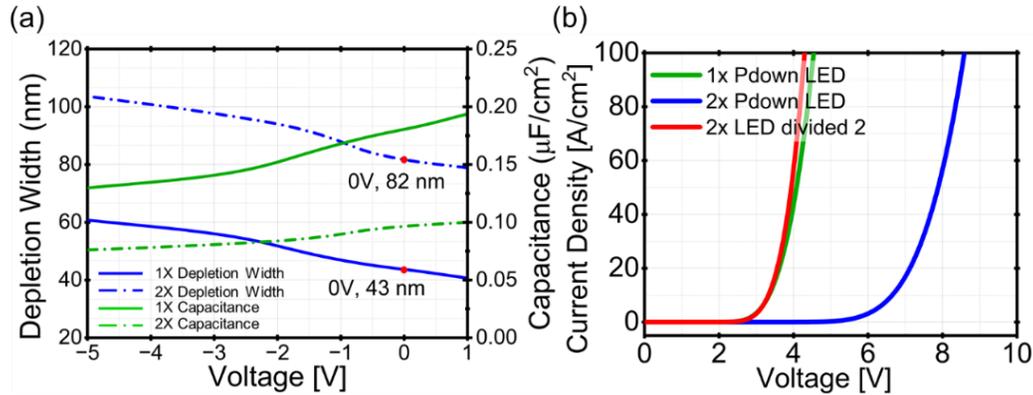

**Figure 3.** Electrical characteristics of the dual junction and single junction p-down LED (a) C-V and (b) J-V.

As expected, the equilibrium depletion width from capacitance-voltage profiling (shown in figure 3(a)) for the single-junction LED (~43 nm) was approximately half that for the double-junction LED (~82 nm) and was in good agreement with the equilibrium energy band diagram (not shown here). Electrical characterization of the dual junction LEDs shows excellent forward voltage scaling (figure 3(b)). At 20 A/cm², forward voltage for single junction LED was measured to be 3.6 V and 7.1 V for dual junction. The double-junction device has a voltage drop that is very close to twice that of the one-junction device (Figure 2 (b)). The operating voltage of the LEDs (~3.6 V/junction) is still degraded by the excess voltage in the tunnel junction. Since similar ohmic metal electrodes (top and bottom) are formed to n-type layer with similar doping/thickness for single junction and dual junction LED and the measurement is performed on devices with similar dimensions, the effect of electrodes on the electrical performances can be assumed similar for both the device types. We attribute the excess voltage in each single p-down LED+TJ structure to the tunnel junction – MOCVD-grown tunnel junctions in the $n^{++}/p^{++}$ configuration have high voltage drop due to challenges associated with achieving high Mg doping density in the top p-region while ensuring that the active region does not degrade due to the Mg doping tail. The estimated tunnel voltage drop in these devices is still the lowest voltage demonstrated for any MOCVD-grown reversed-polarization tunnel junction-based green LED.

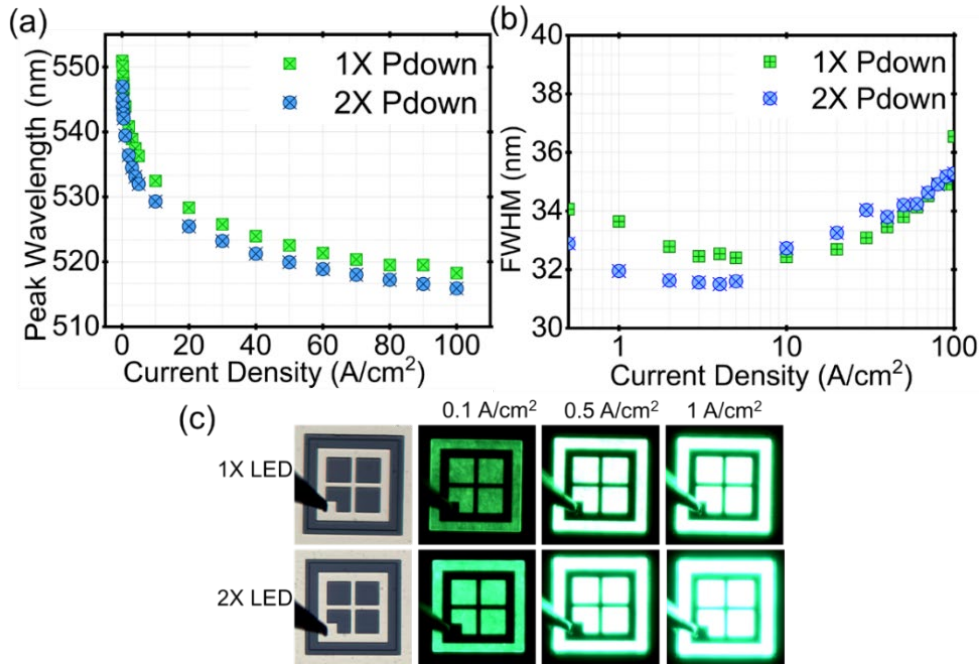

**Figure 4.** (a) Peak wavelength shift with increasing current density from electroluminescence measurement (b) Extracted full-width half maximum (FWHM) from the measured EL spectra (c) Optical micrograph of the devices under low current injection.

    The electroluminescence peak shift extracted from the measured electroluminescence spectra is shown in figure 4(a). Both multi-active region p-down LED and single junction p-down LED show similar wavelength shift from low high current density indicating close wavelength matching between the two active regions. The peak emission wavelength for both multi-active region and single junction LEDs shifts from 543 nm at 0.1 A/cm$^2$ to 517 nm at 100 A/cm$^2$. The full-width half maximum (FWHM) extracted from the electroluminescence spectra for dual and single junction LEDs also show a similar trend with variation of current densities (figure 3(b)) indicating minimal degradation in the active region for dual junction devices. Figure 3(c) shows the optical micrograph of the devices under operation at low current density. The devices show uniform emission at low current densities (0.1 A/cm$^2$) for both single junction and dual junction indicating uniform Mg activation in the buried p-type layers and minimal degradation in the active region for dual junction devices.

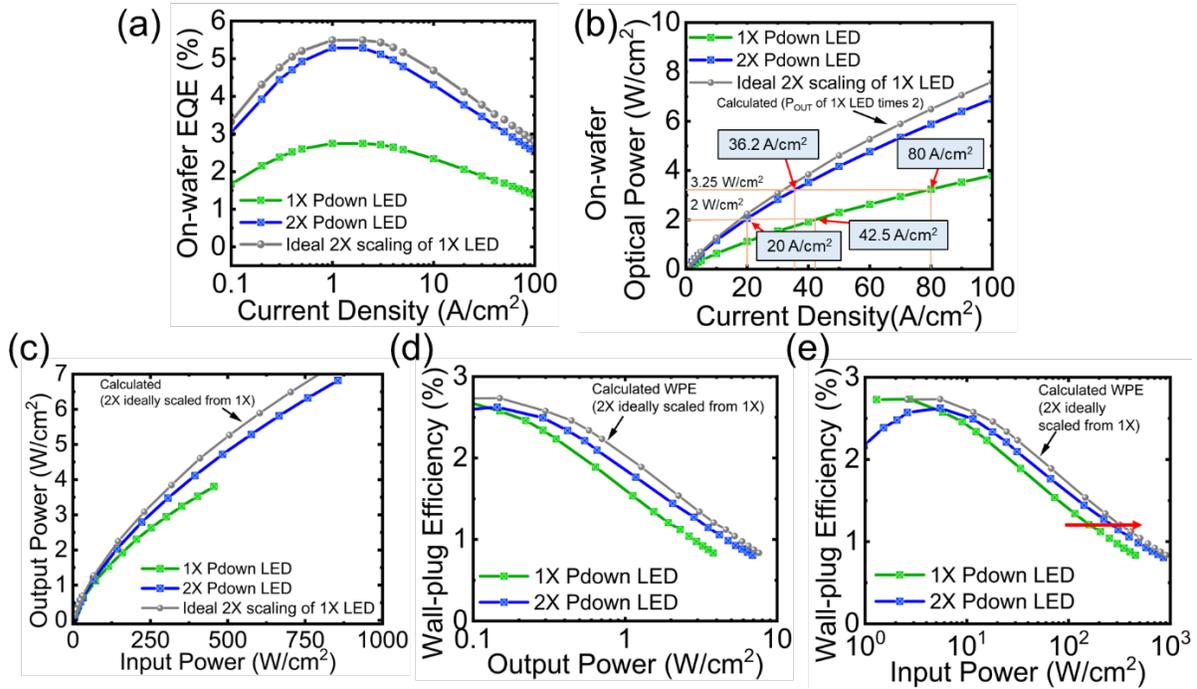

**Figure 5.** (a) On-wafer measured EQE for single junction and dual junction devices. (b) On-wafer measured optical power. (c) On-wafer electrical input power and output optical power for single junction and dual junction LED (d), (e) Calculated wall-plug efficiency from on-wafer electrical and optical measurements for single junction and dual junction LED with their corresponding output power and input power respectively.

On-wafer external quantum efficiency (EQE) measurements are shown in figure 5(a). The single junction LED exhibits peak EQE of 2.75% at a low current density of 1 A/cm$^2$. Dual junction LED shows ~1.9 times scaling of peak EQE, reaching 5.3% at similar low current density of 1 A/cm$^2$. At a high current density of 100 A/cm$^2$, the single junction LED displays EQE of 1.4%, while the dual junction shows an EQE of 2.5%. While the scaling of EQE is not exactly 2×, the significant improvement of the EQE value both at low and high current densities highlights the benefits of cascading multiple active region to tackle pronounced efficiency droop in long wavelength emitters. Several factors such as the increased thermal budget and Mg memory effects can impact the optical characteristics of multi-active region LEDs and cause non-ideal EQE scaling.[27, 28] On-wafer measured optical power (figure 5(b)) demonstrates ~1.9× optical power scaling at low current densities of 1 A/cm$^2$ which shifts to around ~1.8× power scaling at 100 A/cm$^2$. The data demonstrates how the optical power scaling allows for significant reduction in the current drive requirement for dual junction LEDs when compared to single junction LEDs and such reduction of current drive for dual junction leads to improvement in the wall-plug efficiency of the device. For the dual active region LED, the improvement of the wall-plug efficiency is achieved from the significantly reduced output forward current, $I_F$ requirement for any given radiant output power. Figure 5(b) shows the on-wafer output optical power comparison with increasing current density for single junction and dual junction LED. As indicated in the figure 5(b) and table 1, for an output optical power of 2 W/cm$^2$, single junction LED requires a forward current, $I_F$ of 42.5 A/cm$^2$, whereas a similar dual junction LED provides 2 W/cm² output power at a forward current density, $I_F$ of 20 A/cm² which is less than half the current requirement from single junction LED. The current requirement further reduces as we move towards higher optical output power. As shown in the figure 5(b), for an output power of 3.25 W/cm², a single junction LED requires $I_F = 80$ A/cm² whereas the dual junction requires $I_F = 36.2$ A/cm². Such reduced $I_F$ requirement allows the LED to operate at higher radiative efficiency regime with higher optical output power compared to single junction LED. Figure 5(c) shows the output power of single junction and dual junction LED with increasing input electrical power. Figure 5(d) and figure 5(e) shows the WPE comparison for the two LED with respect to their corresponding output and input power respectively. These show that the dual junction LED is capable of operating at higher wall-plug efficiency for higher input/output power proving their efficacy as an efficient high-power lighting source. Further improvement on the wall-plug efficiency could be achieved in the future with better scaling of the optical power output for multi-active region LEDs. In addition, the lower current requirement also leads to lower wavelength shift for the two-junction LED allowing the device to operate a longer wavelength compared to the single junction LED.

**Table 1.** Performance of single and dual junction p-down green LED at 2 W/cm² on-wafer output power

| Sample | On-wafer output power | WPE (%) | Input power (W/cm²) | Voltage drop (V) | Current density (A/cm²) | Wavelength ($\lambda$) | EQE (%) | EQE /junction (%) |
|---|---|---|---|---|---|---|---|---|
| 1× LED | 2 W/cm² | 1.17 | 170 | 4.0 | 42.5 | 524 | 1.75 | 1.75 |
| 2× LED | 2 W/cm² | 1.40 | 142.4 | 7.12 | 20 | 530 | 3.77 | 1.89 |

In conclusion, this work shows promising results for the potential utilizing tunnel junction-enabled multi-active region LED structures to circumvent efficiency droop phenomenon in long-wavelength visible emitters. We show Ga-polar multi-active region p-down green LEDs with near ideal voltage scaling, high external quantum efficiencies with ~1.9× EQE scaling, and significantly higher wall-plug efficiency than an equivalent single-junction LED. The integration of tunnel junctions can provide a promising path toward high efficiency long wavelength InGaN LEDs along with innovative device designs and applications on reverse-polarity structures.

**Acknowledgements**

This material is based upon the work supported by the U.S. Department of Energy's Office of Energy Efficiency and Renewable Energy (EERE) under the Building Technologies Office award no. DE-EE0009163. The views expressed in the article do not necessarily represent the views of the U.S. Department or the U.S. Government.